\documentclass[noshowpacs,twocolumn,preprintnumbers,secnumarabic,amsmath,amssymb,nofootinbib]{revtex4-1}
\usepackage{dcolumn}      
\usepackage{bm}           
\usepackage{graphicx}

\usepackage{amsmath,amssymb}  
\usepackage{bm}  

\usepackage[pdftex,bookmarks,colorlinks,breaklinks]{hyperref}  
\usepackage{mathrsfs}
\usepackage{amsmath}
\usepackage{bm}
\newcommand{\be}{\begin{equation}}
\newcommand{\ee}{\end{equation}}
\newcommand{\ba}{\begin{eqnarray}}
\newcommand{\ea}{\end{eqnarray}}
\newcommand{\bs}{\begin{subequations}}
\newcommand{\es}{\end{subequations}}

\newcommand{\forget}[1]{\iffalse#1\fi}
\newcommand{\forgetmenot}[1]{\iftrue#1\fi}

\newcommand{\del}{\nabla}

\renewcommand{\div}{\hskip0.9pt{\mathsf{div}\hskip2pt}}

\newcommand{\curl}{\hskip0.9pt{\mathsf{curl}\hskip2pt}}

\newcommand{\dis}{\hskip0.9pt{\mathsf{dis}\hskip2pt}}

\newcommand{\Qs}{Q^{(S)}}
\newcommand{\Qv}{Q^{(V)}}

\newcommand{\Qt}{Q^{(T)}}

\renewcommand{\>}{\rangle}

\newcommand{\V}[1]{{\bm{#1}}}
\newcommand{\T}[1]{{\bm #1}}

\begin{document}
\title{Magnetic fields from second-order interactions}
\author{Bob Osano}
\email{bob.osano@uct.ac.za} 
\affiliation{Astrophysics, Cosmology and Gravity Centre, Department of Mathematics and Applied Mathematics, University of Cape Town, Rondebosch 7701, Cape Town, South Africa}

\begin{abstract}
It is well known that when two types of perturbations interact in cosmological perturbation theory, the interaction may lead to the generation of a third type. In this article we discuss the generation of magnetic fields from such interactions. We determine conditions under which the interaction of a first-order magnetic field with a first-order scalar-or vector-, or tensor-perturbations would lead to the generation of second order magnetic field. The analysis is done in a covariant-index-free approach, but could be done in the standard covariant indexed-approach.\end{abstract}
\date{\today}
\maketitle
\section{{{Introduction}} }
Recent observations seem to show that magnetic fields are ubiquitous in the universe. The extent of the existence of these fields ranges from the core regions to the intra-cluster medium of galaxy clusters (See \cite{Davis:1999bt} and accompanying references). Results of studies of Lyman-$ \alpha$ in high red-shift absorption systems \cite{Lymann} suggests that magnetic fields, of considerable strengths, may be present during the condensations period.This is an intriguing possibility given that it is still unclear when magnetic fields first appeared \cite{Widrowetal:2011}or what role if any they played in structure formation \cite{Kim:1996, Sub:1998, Jed:1998} or the evolution of structures. The origin of such fields is still unresolved.

The study of magneto-genesis is partly driven by the need to explain large-scale galactic fields. Spiral galaxies, for example, may have magnetic fields of the order of a few $\mu$G.These fields may have originated from a comparatively larger primordial seed fields (coming from phase transition or Inflation in the early universe), which may have subsequently been amplified by non linear interactions associated with the collapse of the protogalaxy, or from a weaker one that was then made stronger by the galactic dynamo. However, there are problems with these ways of generating magnetic fields from the early universe: phase-transition mechanism suffers from small correlation length scales, in particular even if a large portion of the energy density of the universe went into magnetic fields, the averaged field would be insignificantly small (unless helicity came into play \cite{Brandenburg:1996, Field&Carroll:2000}). Although Inflation may generate the required correlation length scales, one needs a mechanism for breaking conformal invariance\cite{Widrow:2002}. This means that the seeds fields would have to be amplified. Depending on the efficiency of the dynamo mechanism, seeds ranging from $10^{-30}$G to $10^{-8}$G are required \cite{Tsagas&Maartens:2000}. 

Besides primordial mechanism, there other ways of generating seed fields for example it is known that thermal effects or the Biermann battery effects. For such fields, it is proposed that there are five key stages that the fields go through from the protogalactic stage, namely; Biermann battery, Kinematics dynamo, Magnetized dynamo, inverse cascading and compression, and finally the galactic $\alpha-w$ dynamo \cite{Malyshkin}, each stage requiring a different threshold in order to operate.There is still no way of determining whether the galactic and cluster magnetic fields is from primordial or post-recombination mechanism, given that such histories would be lost due to strong amplification by some of the stages mentioned above; in particular the magnetized dynamo stage which is able to amplify a seed by a factor of $10^{9}$ \cite{Malyshkin}. Our interest in this article is on the generation of magnetic fields by second order effects and conditions that govern the generation of such fields. These fields may then be amplified by mechanisms such as those referred to above. We present our equation, and discussions in the index-free notation introduced in \cite{Clarkson&Osano:2011}.\\

\section{ Irreducible parts and Index-free notation} Following \cite{Clarkson&Osano:2011}, we assume a background that has FLRW geometry of curvature $K$, and upon which all relations are defined. The following commutation relationships will hold for objects of perturbative order $m$, but curvature corrections have to be introduced for objects of lower perturbative orders. It is also assumed that $c=1$ and $8\pi G=1$ and the existence of a 4-velocity that will be denoted by $u^{a}$. It the possible to define the spatial metric $h_{ab}=g_{ab}+u_au_b$ which projects orthogonal to the 4-velocity. As in \cite{Clarkson&Osano:2011}, all rank-1 and -2 tensors are orthogonal to the 4-velocity, in addition all rank-2 tensors are symmetric and trace-free. Angle brackets on indices are used to indicate this. The conformal spatial covariant derivative acting on scalars or spatial tensors is defined as $\overline{\nabla}_f= a ( h_f^{~b}\del_b)=a(D_f$), where the coefficient $a$ is the scale factor and $D_f$ is the spatial derivative used in the covariant approach. We use $\overline{\nabla}_f$ as it commutes with $u^f\overline{\nabla}_f$ and is the covariant derivative normally used in the metric approach to perturbation theory. The irreducible parts of the spatial derivative of PSTF tensors are the divergence ($\div T_{b\dots c}= \overline{\nabla}_a T_{ab\dots c}$), the curl: ($\curl T_{ab\dots c}=\varepsilon_{de\<a}\overline{\nabla}^d T_{b\dots c\>}^{~~~~~e}$), and the distortion: ($\dis T_{ca\dots b}=\overline{\nabla}_{\<c}T_{a\dots b\>}.$) Note that the divergence decreases the rank of the tensor by one, the curl preserves it, while the distortion increases it by one. 
Index free notation may be used for any equations which are irreducibly split, and all objects appearing are similarly split. In this case we denote a 3-vector $V^a$ by $\V{V}$ and, more generally, a PSTF tensor $X_{a\cdots b}$ by $\T{X}$. We define three products between vectors and PSTF tensors. For this, let $\V{V}, \V{W}$ be rank-1, $\T{X}, \T{Y}$ be rank-2 and $\T{Z}$ be rank-3 (rank-3 objects only commonly appear as distortions of rank-2 tensors).\\{\it Dot products} :($ 
 V^aW_a = \V{V}\cdot\V{W},V^b X_{ab}=\V{V}\cdot\T{X},X^{ab}Y_{ab}=\T{X}\cdot\T{Y}, Z_{abc}X^{bc}, \T{X}\cdot\T{Z}, Z_{abc}Z^{abc},\T{Z}\cdot\T{Z}, Z_{abc}X^{bc}, \T{Z}\cdot\T{X}, Z_{abc}V^{c}=\T{Z}\cdot\V{V}$)
 \\{\it Cross-products:} ($ \epsilon_{abc}V^bW^c=\V{V}\times\V{W},\epsilon_{abc}{{X}^{b}}_{d}Y^{cd}=\T{X}\times\T{Y},\epsilon_{abc}{Z^{b}}_{ef}Z^{cef}=\T{Z}\times\T{Z}$) \\{\it Circle products}:($V_{\langle a}W_{b\rangle}=\T{V}\circ\T{W},{X_{\langle a}}^{c}Y_{b\rangle c}=\T{X}\circ\T{Y},Z_{bc\langle a}{Z^{bc}}_{d\rangle}=\T{Z}\circ\T{Z}$) One can considerably simplify the appearance of the covariant equations by expressing any product using these index-free notation. We refer the reader to \cite{Clarkson&Osano:2011}, for a detailed discussion.\\

\section{Splitting in the covariant perturbation theory.} Techniques used perturbation theory in cosmology are based on the decomposition of perturbations into scalar, vector and tensor parts, on a constant-curvature background. This split is but a generalization of HelmholtzÕs theorem that has been extended to include tensors on 3-spaces; having constant curvature\cite{Stewart&Walker:1974, Stewart:1990, Kodama&Sasaki:1984}. The split is performed non-locally: either harmonically or by integrating over Green's function. This then requires the specification of boundary conditions in order to provide a definition of these non-local variables. We also need to know a perturbation variable everywhere, for us to be in a position to specify a scalar vector or tensor type of perturbation anywhere. For HelmhotzÕs theorem in 3-dimensional flat space, any vector $\V{N}$ can be written in terms of a scalar $S$ and vector $\V{V}$:
$\V{N}= \overline{\nabla}S + curl \V{V}$, where $\overline{\nabla}^2 S = \div \V{N}$ , and $\overline{\nabla}^{2}\V{V} = -\curl \V{N}$,
which comes from the vector calculus identities $ \div\curl \V{N} = 0$ and $\curl\curl\V{N} = -\overline{\nabla}^{2}\V{N} + \overline{\nabla}\div\V{N}$ (in
Euclidean space). Solutions for $S$ and $\V{V}$ can be found if they tend to zero at infinity, which means that both $S$ and $\V{V}$ are essentially non-local, and require the knowledge of $\V{N}$ everywhere in order to be able to give it at a particular point. On the other hand, we can think of $\div\V{N}$ as a pure scalar degree of freedom which is deÞned locally whenever $\V{N}$ is; similarly, $\curl\V{N}$ is a pure vector degree of freedom. So, given $\V{N}$ we can isolate unique, scalar and vector degrees of freedom which are locally defined. This will be of use when considering perturbation modes that interact with linear-order magnetic fields. Although, non trivial, it is possible to recover non-local equivalent of the various degrees of freedom \cite{Clarkson&Osano:2011}. In particular $\dis\curl\div\T{\sigma}$ and $\curl\div\V{N}$ would yield the vector parts of rank-1 and rank-2 tensors respectively 

We will adopt the covariant theory (see \cite{Ellis&Elst:1998} and the references there in for a comprehensive review), subject to the above splitting, our study. We take the following as give: (1) there exists a vector field $u^a$ that is time like ( $u^a u_a= -1$), (2) based on the 4-velocity one can define a projection tensors ${h_{a}}^{b}={\delta_{a}}^{b}+u_{a}u^{b}$ such that ${h_{a}}^{b}u^{b}=0,$ where ${\delta_{a}}^{b}$ is the tensor used in the metric approach. (3) With the help of the 4-velocity, tensor objects may be split into their invariant Ôtemporal' and Ôspatial parts. These parts are scalars, 3-vectors, and projected, symmetric, trace-free (PSTF) tensors. (4) We focus on perfect fluids, but note the analysis could be performed for other fluids. (5) We also only consider a background in which the $^3 R$ scalar vanishes( i.e. leading to $\mu=\frac{1}{3}\Theta^2$), again this is just to simplify the presentation and one can consider a more general case.
\section{Gauge-invariance} We want to study magnetic fields that are induced by interactions at second order and will use a perturbative scheme within the covariant theory. It is therefore necessary to find away to separate the first order and the second order fields in a gauge-invariant manner. To this end, we are guided by the Stewart-Walker lemma \cite{Stewart&Walker:1974}. A gauge invariant quantity at a particular order will be that quantity that vanishes or is constant at all lower orders. Since our background is FLRW, the magnetic fields and all perturbations are first order or higher in our perturbative scheme. There are three ways of formulating a gauge-invariant second order magnetic fields: 

{\it Method 1:} We take the magnetic field as purely of first order, the Maxwell equations represent the fields that are induced when the first order fields interact with gravitational perturbations. We can then use the notations $\V{B}^{(1)}$ and $\V{B}^{(2)}$ for first and second order fields respectively. This would be based on the splitting of the total magnetic field as follows: $\V{B}=\V{B}^{(1)}+\V{B}^{(2)}$, where $\V{B}^{(2)}$ is taken to be a gauge-invariant second order term from the interaction. A form of this splitting was found to be inconsistent in \cite{Gary&Zunkel}, an inconsistency that emanated from a commutation relation involving the time and  thespatial derivative ($D$) of a vector. In our case the time and the spatial derivative involving $\overline{\nabla}$ commute and the inconsistency does arise. 

{\it Method 2:} One can define a second order term based on an equation that vanishes at the first order. The propagation equation for first order field is given as $\dot{\V{B}}^{(1)} + 2H\V{B}^{(1)} =0,$ one can then define the second order variable $\V{\beta}=\dot{\V{B}}^{(1)} + 2H\V{B}^{(1)}$ as was done in \cite{Gary&Zunkel, Bishop:2012}. 

{\it Method 3:} The third approach is to define a variable related to the first order field, but which vanishes at first order, for example $\V{\beta}=\curl\overline{\nabla}\div\V{B}$ vanishes first order given that $\div\V{B}=0$. We have then constructed a second order object by taking the gradient of the rank-0 tensor. The $\curl $ then eliminates the scalar part of this object leaving only the vector part ( see \cite{Clarkson&Osano:2011} for an elaborate discussion about how to construct vectors from scalars). We are now at a point where we can discuss magnetic induction. We only give detailed discussions of methods 1 and 3 (method 2 can be found in \cite{Gary&Zunkel, Bishop:2012})

\section{ Maxwell's equations.}
Assume that at first order the electric fields are small compared to the magnetic fields ($\V{E}^2\ll \V{B}^2$). At linear-order \ba \label {eq:ind1}\dot{\V{B}}^{(1)} + 2H\V{B}^{(1)} =0, \ea with $\div\V{ B}^{(1)}=0$, where $H=\Theta/3$. (2) The vorticity is taken to be zero (a restriction which could be relaxed, but vorticity is not easily generated see for example \cite{Tsu:2007}). The coupled Einstein-Maxwell equations up to second order are then given by (\cite{Tsagas&Barrow:1997},\cite{Ellis&Elst:1998} and \cite{Marklundetal:2003}).
\begin{eqnarray}
\label{eq:induction_B}
\dot{\V{B}}^{(2)} + 2H\V{B}^{(2)} &=& \V{\sigma}^{(1)}\cdot {\V{B}}^{(1)} - (\frac{\overline{\nabla}}{a}\times\V{E} )^{(2)}
\\
\label{eq:induction_E}
\dot{\V{E}} ^{(2)}+2H\V{E} ^{(2)}&=&(\frac{\overline{\nabla}}{a}\times\V{B})^{(2)},
\end{eqnarray} where superscript 1 denotes objects of perturbative $\mathcal{O}(1)$ while superscript 2 denotes objects of perturbative order $\mathcal{O}(2)$. The superscripts represent a conceptual separation and not an actual physical separation of the orders.  We have dropped products of order $\mathcal{O}(3)$ and higher. We have assumed that the vorticity vanishes ($\V{\omega}=0$) and a quasi-neutrality approximation, i.e. the total charge density satisfies $\rho\approx 0$, which is a viable assumption when electrons and ions follow the same motion.

In the covariant approach to perturbation theory, the solutions of perturbed differential operators are never sought. One can get around this by making sure that the differential operators involved operate on quantities of the corresponding perturbative order.\\

\section{Magnetic wave equation based on method 1} By taking the time derivative of equation (\ref{eq:induction_B}) and using  equation (\ref{eq:induction_E}), equation (\ref{eq:ind1}) and the time derivative for the shear tensor up to first order, one finds induction equation
 \begin{eqnarray}
\label{Induct}&\ddot{\V{B}}^{(2)}&+\frac{1}{a^2}\overline{\nabla}\times\overline{\nabla}\times\V{B}^{(2)}+5H\dot{\V{B}}^{(2)}+\left[3(1-w)H^{2}\right]\V{B}^{(2)}\nonumber\\\ &=&\dot{{\T{\sigma}}}\cdot{\V{B}^{(1)}}+H{\T{\sigma}}^{(1)}\cdot\V{B}^{(1)},
\end{eqnarray} 
The source term is made up rank-1 tensor ${\T{\sigma}}\cdot\V{B}=\V{I}$ and its time derivative ( note that the first term $\dot{{\T{\sigma}}}\cdot{\V{B}}$ can be written as $\dot{\V{I}}+2H\V{I}$, thus the source is generated by the quasi-vector term $\V{I}$. The time derivative would not change the rank of the vector object. This object is no longer a pure vector as it has both a scalar and a vector part. As a consequence, the induced field is not a pure magnetic field ( i.e. $\div\V{B}^{(2) }\neq 0$). In order to examine magnetic fields at second order, we need to extract the pure vector part for this equation. Pure vector part is given by $\curl \V{I}$ (i.e. $\div\curl\V{I}=0$), since the curl would set the scalar part to zero \cite{Clarkson&Osano:2011}. The correct induction equation for magnetic fields up to second order is then given by 
 \begin{eqnarray}
\ddot{\V{\mathcal{B}}}+\frac{1}{a^2}\overline{\nabla}\times\overline{\nabla}\times\V{\mathcal{B}}+5H\dot{\V{\mathcal{B}}}+\left[3(1-w)H^{2}\right]\V{\mathcal{B}}= \mathcal{S},
\end{eqnarray}
where\begin{eqnarray}
\mathcal{S}=\curl(\dot{{\T{\sigma}}}\cdot{\V{B}^{(1)}})+H\curl({\T{\sigma}}^{(1)}\cdot\V{B}^{(1)}),\end{eqnarray} where $\V{\mathcal{B}}=\curl \V{B}^{(2)}.$ and $\V{\mathcal{S}}=\curl {S}$. We emphasize that in order for one to study the fields that are induced when linear order field interact with other linear order modes of perturbation, one aught to examine conditions which may lead to the to $\div(\curl \V{I})=0$. It is a requirement that the induced $\V{B}^{(2) }$ field be a pure vector, for it to be understood as pure magnetic field. This requirement is at the heart of our analysis and deserves detailed consideration. This requirement will be achieved if $\curl \V{I}^{a}$ is divergence-free (note that $\V{I}$ generates the source to the induced field). We discover the restrictions on the shear tensor which then forms the basis for all subsequent analysis. First note that in the standard covariant notations,
 $\overline{\nabla}_{a}\curl\V{I}^{a}=\overline{\nabla}_{a}\curl(\T{\sigma}^{ab}\V{B}_{b}^{(1)})$ which in the index-free form is $\div\curl(\T{\sigma}\cdot\V{B}^{(1)})$.  It can be shown that \begin{eqnarray}
\label{GW1}
\div\curl(\T{\sigma}\cdot\V{B}^{(1)})
&=&-\frac{1}{2}\div(\T{\sigma}\cdot\curl\V{B}^{(1)})+\div(\V{B}\cdot\curl\T{\sigma})\nonumber\\
&&~~~~~~~~~~~-\div(\T{\sigma}\times\dis\T{B}),
\end{eqnarray}\\ 
\begin{eqnarray}
\curl \left(\V{V}\cdot {\T{X}}\right)
&=& -\frac{1}{2}{\T{X}}\cdot\curl \V{V}  + \V{V} \cdot \curl
{\T{X}} -  {\T{X}}\times \dis \V{V},\nonumber\\ 
\end{eqnarray} where we have used the identity derived in \cite{Clarkson&Osano:2011}. The terms in equation (\ref{GW1}) can separately be expanded to give: 
 \begin{eqnarray}
\label{s2}\div(\V{B}\cdot\curl\T{\sigma})&=&\curl\T{\sigma}\cdot\dis\V{B}+\V{B}\cdot\div\curl\T{\sigma},\\
\label{s3}-\div(\T{\sigma}\times\dis\T{B})&=&\T{\sigma}\cdot\curl\dis\V{B}-\dis\V{B}\cdot\curl\T{\sigma},\\
\label{s1}-\frac{1}{2}\div(\T{\sigma}\cdot\curl\V{B}))&=&-\frac{1}{2}\T{\sigma}\cdot\dis\curl\V{B}-\frac{1}{2}\curl\V{B}\cdot\div\T{\sigma},\nonumber\\
\end{eqnarray}  where the following identities have been used
\begin{eqnarray}
\div \left(\V{V}\cdot {\T{X}}\right)&=& {\T{X}}\cdot\dis \V{V} + \V{V}\cdot\div {\T{X}},\\
\div\left(\T{X}\times\T{Y}\right)&=& -\T{X}\cdot\curl\T{Y}+\T{Y}\cdot\curl\T{X},
\end{eqnarray} where $\T{X}$ and $\T{Y}$ are tensors of rank 2, and $\V{V}$ is a tensor of rank 1.
The first term on the RHS of equation (\ref{s2}) will cancel the second term on the RHS of equation (\ref{s3}). Using the identity $\dis\curl=2\curl\dis$\cite{Clarkson&Osano:2011} the first term on the RHS of equation (\ref{s1}) takes a form similar to the first term of equation (\ref{s3}) but with the opposite sign and should therefore cancel. Only the second terms on the RHS of equations (\ref{s1}) and (\ref{s2}) will be left: \[\V{B}\cdot\div\curl\T{\sigma}-\frac{1}{2}\curl\V{B}\cdot\div\T{\sigma}.\] 
 Now consider the $\curl$ of the zero-rank tensor (..or a scalar). It follows that\[
0=\curl(\V{B}\cdot\div\T{\sigma})=\curl\V{B}\cdot\div\T{\sigma}+\V{B}\cdot\curl\div\T{\sigma},
\] and hence $
-\frac{1}{2}\curl\V{B}\cdot\div\T{\sigma}=\frac{1}{2}\V{B}\cdot\curl\div\T{\sigma}.
$ This means $\div\curl\V{I}=\frac{3}{2}(\V{B}\cdot\curl\div\T{\sigma})=3(\V{B}\cdot\div\curl\T{\sigma}).$ 
The last equality comes form the identity $\curl\div=2\div\curl\label{curldiv-t}$ \cite{Clarkson&Osano:2011}. It is clear that $\div\div\curl\V{I}=3\div(\V{B}\cdot\div\curl\T{\sigma})=0,$ since dot product yields a tensor object of rank 0. The divergence os such an object does not exist. This implies that if one were to apply the $\div$ operator on the wave equation (\ref{Induct}) both the side would vanish. This would not prove anything other than the fact that we have demonstrated that ($0=0$!). We have not discussed the complete source term $\V{\mathcal{S}}=\dot{\V{I}}+3H\V{I}$ for two reasons: (1) it is generated by $\V{I}$ and (2) the time derivative commutes with the $\overline{\nabla}$ operator and hence the effects of the $\div$ and $\curl$ are carried through time derivative.

We now examine conditions that will cause $\div\curl\V{I}=0$, but which do not lead to the vanishing of both sides of the wave equation.\\

\subsection{ Vector-Tensor Interactions:} We note that $3(\V{B}\cdot\div\curl\T{\sigma})=0$ if $\div\T{\sigma}=0=\div\T{H}$( pure tensors). It is known that a divergence free magnetic part of the Weyl tensor is a condition for the existence of gravitational waves in the covariant theory. The implication here is that magnetic fields are induced at second-order when linear order fields interact with first order tensor perturbations (gravitational waves). Similar fields were studied in \cite{Gary&Zunkel, Tsagas:2002}

\subsection{Vector-Scalar Interactions:}
 The other possibility is $3(\V{B}\cdot\div\curl\T{\sigma})=0,$ if $\curl\T{\sigma}=0.$ The implication is that only the scalar mode of shear is excited, of course $\T{H}=0$. This switches off the gravitational waves. Therefore the condition required for the analysis of fields induced by interaction with density perturbations is $\curl\T{\sigma}=0.$ These fields have recently been studied in \cite{Bishop:2012}.

\subsection{Vector-Vector Interactions:}
vector-vector interaction is a tricky one and has not. So how would we extract the fields that are induced by interaction with vectors? If we consider a situation where only vector modes are excited an linear order, then the residual term would not vanish. This implies that when the linear magnetic fields interact with linear order vectors the resulting wave equation is not that of pure vector and therefore a vector extraction needs to be performed. Such an extractor was developed in \cite{Clarkson&Osano:2011}.

\section{Fourier decomposition:} Inorder to solves the wave equation, one can define the a scalar, vector and tensor harmonic basis in the usual way; $\overline{\nabla}^2\Qs=-k^2\Qs$, for vectors, $\overline{\nabla}^2\Qv_a=-k^2\Qv_a$( where we have two parities of orthogonal harmonics, $(k^2+2K)^{1/2}\Qv=\curl\Qv\Leftrightarrow (k^2+2K)^{1/2}\Qv=\curl\Qv$) and for tensors $\overline{\nabla}^2\Qt_ab=-k^2\Qt$ (the parities are related as follows $(k^2+3K)^{1/2}\Qt=\curl\bar{Q}^{T}\Leftrightarrow (k^2+3K)^{1/2}\bar{Q}^{T}=\curl\Qt$ \cite{Tsagas&Roy}.)The wave equation can appropriately be decomposed and the desired field studied. 
\section{Magnetic wave equation based on method 3} We now consider the wave equation for $\V{\beta} (\equiv\curl\overline{\nabla}\V{B})$. First note that $\div\V{\beta}=\div\curl\overline{\nabla}\div\V{B}=0$ ( see\cite{Clarkson&Osano:2011} for how to extract various kinds of perturbations and  \cite{Chris:2012} for how to construct gauge-invariant quantities in the covariant theory). In order to simplify presentation, we write $\V{\beta}=\mathcal{L}[\V{B}]=0,$ where $\mathcal{L}\equiv \curl\overline{\nabla}\div.$ It can be shown that:
\ba
\dot{\V{\beta}}+2H\V{\beta}=\mathcal{L}[\T{\sigma}\cdot\V{B}-\curl\V{E}].
\ea taking the time derivative of this equation and making use of the propagation equation for $\V{E}$, we get the wave equation

\begin{eqnarray}
\ddot{\V{\beta}}+\frac{1}{a^2}\overline{\nabla}\times\overline{\nabla}\times\V{\beta}+5H\dot{\V{\beta}}+\left[3(1-w)H^{2}\right]\V{\beta}=\V{ \mathcal{S}},
\end{eqnarray} \ba\V{\mathcal{S}}=\mathcal{L}[\dot{\V{I}}]+2H\mathcal{L}[\V{I}] ,\ea where $\V{I}=\V{\sigma}\cdot\V{B}.$ This system is identical to the system in method 1 and the findings in that system hold here as well. The difference is that, in this approach, the second order fields have been formulated in an explicit manner.

\section{Conclusion:} We have examined the wave equation for the $\V{B}$ induced by the interactions of a linear order field with linear order of perturbations. We find that the appropriate wave equation for this purpose is that of $\V{\mathcal{B}}(\equiv\curl \V{B}^{(2)})$ rather than the traditional $\V{B}^{(2)}$. We have also considered cases where specific types of perturbations are excited at first order and have discovered the conditions, which if placed on the shear tensor, would allow the study of various forms of interactions. We found a gauge -invariant  variable $\beta=\curl\overline{\nabla}\div\V{B}$ which encodes second order magnetic fields. We have not discussed the era in which the wave equation, together with the various conditions on the shear, would applicable. This will be presented in a future article.

\appendix*

\end{document}